# Illuminating Fourier Series with Audacity

Amy C. Courtney,[1] and Michael W. Courtney[2]

[1]BTG Research, P.O. Box 62541, Colorado Springs, CO, 80962
Amy_Courtney@post.harvard.edu

[2]United States Air Force Academy, 2354 Fairchild Drive, USAF Academy, CO, 80840
Michael_Courtney@alum.mit.edu

**Abstract:** This paper briefly describes some simple techniques for illuminating Fourier series and Fourier analysis with the Audacity sound recording program. These techniques are easily applied in the classroom and help students move beyond formula roulette in their understanding of Fourier techniques.

## I. Introduction

The essential ideas underpinning Fourier analysis have applications in a wide variety of areas including radar, sonar, communications, sound signature determination, optical pattern recognition, signal processing, etc. However, these essential applications are usually considerably removed from the sterile formulas of the mathematics courses where Fourier series and Fourier analysis are introduced. This paper briefly describes some simple techniques that empower student learning by adding both auditory and visual representations of the underlying phenomena and mathematics.

Audacity is a free, open source audio program (Sendall et al., 2008; http://audacity.sourceforge.net/) for recording, editing, and analyzing sounds. A number of platforms and operating systems are supported, including Mac OS X, Microsoft Windows, and Gnu/Linux. Although it supports more sophisticated applications, here it is used with the computer's built in microphone to record sounds, display their waveforms, and compute their Fourier transforms to reliably reveal frequency components.

In addition to this application, the computer sound digitization capabilities of Audacity have also found classroom and experimental use in determining time of flight (thus average velocity over a known distance) of various projectiles including kicked and dropped balls,(Aguiar and Pereira, 2009; Courtney and Courtney, 2011) potatoes launched from a potato cannon,(Courtney and Courtney, 2007) and rifle bullets, (Courtney, 2008; Courtney and Courtney, 2009) as well as in direct measurement of the speed of sound by measuring the time delay from the sound of a balloon popping to it's return echo from the side of a building a known distance away.(Berg and Courtney, 2009) Audacity has also found scientific application in recording and analyzing sounds from a variety of natural sources (Hockicko and Jurečka, 2007; Tallechea et al., 2011).

## II. Method and Results

Once the software is downloaded and installed on a computer with a built-in microphone, its use is intuitive. The user interface has buttons for record, pause, rewind, play, stop, etc. When the sound source is ready, the user just hits the record button. The waveform in Figure 1 is the sound of one of the authors (MC) whistling a low, steady tone.

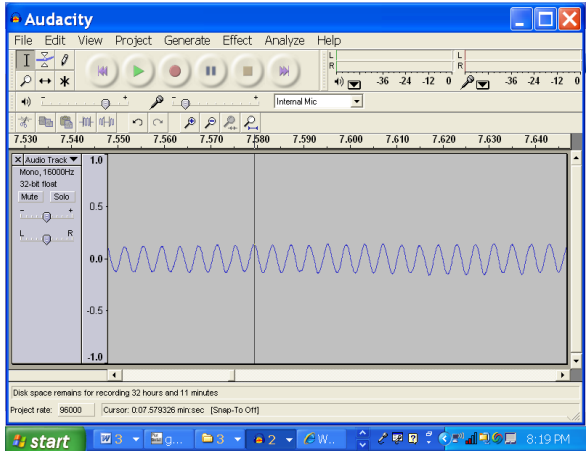

*Figure 1: A sound waveform of whistling a low, steady tone can look a lot like a pure sine wave.*

This wave can be modeled as a function

$f(t) = A + B \cos(Ct - D)$,

where A is the offset, B is the amplitude, C is 2pi/period, and D is the horizontal shift of the sinusoidal function. Modeling the function in this way, or at least determining the period and frequency by inspection is useful for confirming the frequency determination in the following Fourier analysis. Using the cursor shows that the period (time between successive peaks) is 3.906 ms, which corresponds to a frequency of 256 Hz, a musical "C" note.

The Fourier transform of the waveform reveals its frequency components and can be computed by highlighting a region of the waveform and selecting the menu item "Analyze -> Plot Spectrum" which produces the result shown in Figure 2 below.

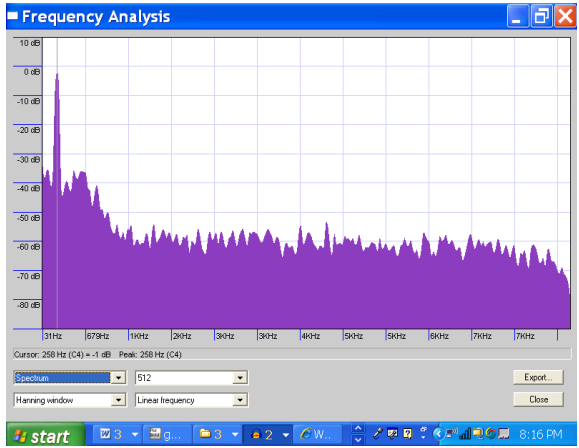

*Figure 2: Spectrum of whistled low tone is dominated by a single peak at 258 Hz. Note the logarithmic scale (dB). The next highest frequencies are down by 30 dB relative to the peak, indicating amplitudes roughly 1000 times smaller.*

Using the cursor reveals that the tallest peak is located at 258 Hz. The small difference between the time and frequency domains might be due to an inaccuracy introduced by the short time window or by the changing frequency of an imperfect whistler.

Figure 3 shows the sampled sound waveform of the high D note played on a child's recorder. The waveform is obviously more complex, but using the cursor, the period of two highest adjacent peaks can be determined to be 0.809 ms, corresponding to a frequency of 1236 Hz.

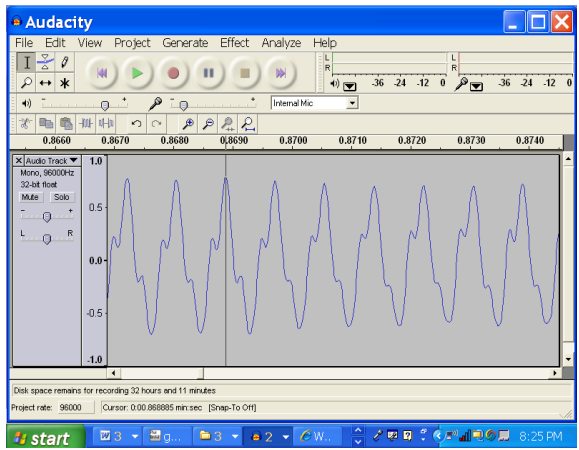

*Figure 3: Sound waveform of a recorder.*

Highlighting a section of the waveform 0.2s long and selecting "Analyze -> Plot Spectrum" produces the graph shown in Figure 4. Using the cursor reveals the tallest peak at 1205 Hz, corresponding to a D note, but not exactly the same frequency as determined from the waveform. This is likely due to the recorder failing to hold the same frequency over the sampling interval. Navigating to subsequent peaks with the cursor shows a small peak at 2400 Hz (the second harmonic), a large peak at 3592 Hz (the third harmonic), as well as additional peaks at 4807 Hz, 6002 Hz, 7182 Hz, and 8388 Hz (the $4^{th}$ – $7^{th}$ harmonics). Thus, although the sound waveform is dominated by the $1^{st}$ and $3^{rd}$ terms in the Fourier series, the recorded waveform gives evidence of terms out to the $7^{th}$ harmonic.

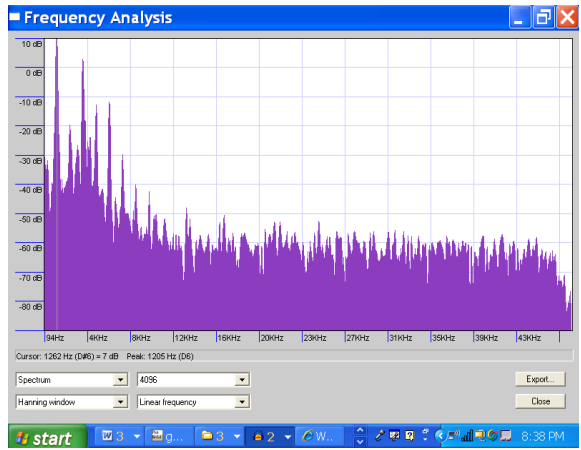

*Figure 4: Spectrum of recorder playing D note. Frequencies corresponding to several harmonics are also easily identified. This spectrum is dominated by the first and third terms in the Fourier series, and terms out to the seventh are visible in the spectrum.*

Students often enjoy sampling sounds from their own musical instruments or from simple resonant objects like a wine glass or cymbal. The occasional brave student will volunteer to sing a tone to allow harmonic analysis of the human voice. Audacity also has a spectrogram mode that displays a time-dependent frequency analysis (Select the upside down triangle to the right of "Audio Track" and choose the "spectrum" display mode.) One can record someone talking or singing and then view the spectrogram as the cursor moves along while the recording is replayed. Advanced courses can sample a well-tuned drum head and obtain the Fourier series of Bessel functions. (In practice, this is tricky and one needs to have some skill in tuning drum heads for even tension over the surface for this to work successfully. Alternatively, if one knows what the Bessel series should look like, one can sample in spectrum mode and tweak the tension knobs in real time until the spectrum agrees with expectations.)

### III. Conclusion and Discussion
It is not uncommon for this activity to be a student favorite and the best-remembered lesson of the semester. Depending on the interests of the students, the conversation and activity can be steered in a variety of relevant directions and/or a follow-up activity can be planned where students have a chance to implement new ideas that result from brainstorming and interest generated in the first session. Two frequency sources offset by 1-3 Hz can be used to demonstrate beats. The difference between nasal, gravelly voices (Bob Dylan) and crystal clear voices (Whitney Houston) and between highly over-toned instruments (violin) and relatively pure instruments (flute) can be seen as well as heard. One can also demonstrate spectral signatures of different coins and brass cartridge cases hitting the ground.

This activity uniquely combines a hands-on approach that includes auditory and graphical representations to the fundamentals of Fourier series and Fourier analysis that is a positive supplement to the conventional theorem and problem-solving introduction to Fourier series. (Courtney and Althausen, 2006)